\documentclass[aps,prl,twocolumn,groupedaddress,showpacs]{revtex4}
\usepackage{graphics}

\begin{document}

\title{Infinite average lifetime of an unstable bright state \\ in the green fluorescent protein}
\author{P. Didier, L. Guidoni, and F. Bardou}

\address{Institut de Physique et Chimie des Mat{\'e}riaux de Strasbourg,\\
UMR 7504 CNRS-ULP-ECPM, 23 rue du Loess,B.P. 43  F-67034
Strasbourg CX, France} \email{ pdidier@ipcms.u-strasbg.fr}


\date{\today}

\begin{abstract}
The time evolution of the fluorescence intensity emitted by
well-defined ensembles of Green Fluorescent Proteins has been
studied by using a standard confocal microscope. In contrast with
previous results obtained in single molecule experiments, the
photo-bleaching of the ensemble is well described by a model based
on L\'evy statistics. Moreover, this simple theoretical model
allows us to obtain information about the energy-scales involved
in the aging process.
\end{abstract}

\maketitle
Since the beginning of the 90's, single molecule experiments have
opened the way to the study of the optical properties of a single
emitter \cite{Moerner99}. In particular, these experiments have
evidenced that an aging mechanism affects fluorescent objects:
each marker performs only a finite number of
absorption/spontaneous emission cycles before definitely reaching
a permanent photo-bleached dark state \cite{Dickson97}. Before
being irreversibly photo-bleached, single emitters often display a
"blinking" behavior: as the time goes on the system undergoes
transitions between a bright state (ON state in which it stays a
time $\tau_{ON}$) and a reversible dark state (OFF state in which
it stays a time $\tau_{OFF}$)\cite{Nirmal96}. In a single molecule
experiment one can measure the light emitted by a single marker as
a function of time and then characterize the blinking and
photo-bleaching behaviors. The study of a multitude of single
systems gives access to the probability distribution associated to
their emission properties, for instance the statistical
distribution of emission spectra \cite{Barkai00,Barkai03} or the
probability distribution $P(\tau_{ON/OFF})$ associated to
$\tau_{ON}$ and $\tau_{OFF}$ \cite{Brokmann03,Margolin05}. In this
way it is possible to obtain more complete information with
respect to an ensemble measurement. Such a strategy works well in
the case of inorganic emitters (i.e. semiconductor quantum dots).
In this case the observation time of a single object (limited by
photo-bleaching) is long enough to obtain a large number of
measurements and therefore meaningful histograms. Moreover, such
systems allow the implementation of parallel measurements on some
hundreds of single emitters (wide field microscopy) opening the
way to study the ergodicity of such systems
\cite{Brokmann03,Margolin05}. Several experiments show that the
$\tau_{ON}$ and $\tau_{OFF}$ are distributed according to broad
laws \cite{Shimizu01,Brokmann03}. Let us stress that a long
observation time is crucial for a complete characterization of
such broad distributions, due to the intrinsic large dispersion of
the measured data. In the case of single organic molecules, the
observation time is strongly reduced by quenching phenomena. This
fast photo-bleaching is a limiting factor for the obtention of
reliable statistical distributions except in special cases of well
controlled environment \cite{Pfab04}. A particularly interesting
case of organic markers is that of fluorescent proteins such as
the Green Fluorescent Protein (GFP) which is one of the most
popular marker for {\it in vivo} imaging in cell biology
\cite{Tsien98}. Single molecule room temperature studies on GFP
show results that are compatible with an aging mechanism governed
by narrow statistical laws (exponential or bi-exponential). With
this assumption it is possible to determine average values such as
$\left<\tau_{ON}\right>$, $\left<\tau_{OFF}\right>$ and the
characteristic time associated with the irreversible transition
between the fluorescent and the dark state (the average
photo-bleaching time) \cite{Peterman99}. Reported average
photo-bleaching times for fluorescent proteins are of the order of
some hundreds of millisecond, depending on excitation intensity
\cite{Harms01}. Therefore single molecule experiments seem to
evidence a fundamental difference between inorganic and organic
emitters: the times associated to aging are distributed according
to broad versus narrow laws. However, the presence of L\'evy
statistics cannot be excluded if we consider the poor quality of
the histograms obtained in the case of proteins.

The question that we address in this letter is : Can one find
signatures of L\'evy statistics in ensemble measurements in the
case of fluorescent proteins? The analysis of the time evolution of
the fluorescence intensity emitted by a well defined ensemble of
GFP under c.-w. excitation shows that this is the case. Starting from
ensemble measurements, we evidence that the GFP
aging process is governed by L\'evy statistics. Moreover,
information about the potential barriers involved in the aging
phenomenon is extracted from the experiment, thanks to a theoretical model that takes
into account an irreversible photo-conversion between a
fluorescent and a permanent dark state. One of the main
results of this study is that the unstable bright state has an
infinite mean lifetime.

In the experiment, conceptually very simple, we have recorded the
time evolution of the fluorescence intensity $F$ emitted by a
well-defined ensemble of a GFP mutant (uv mutant : GFPuv
\cite{Crameri95}) under continuous excitation. The sample is
obtained by dispersing a recombinant GFPuv solution (125~$\mu$M in
a pH=8 buffer solution) in a polyacrylamid matrix \cite{footnote1}. To avoid
denaturation, the protein solution is added after polymerization. The
final concentration in the anhydrous samples is around
$4\times10^{5}$ molecules per $\mu$m$^{3}$. The sample is placed
on a slide-holder of a home-built confocal
microscope illuminated by an Ar-ion laser ($\lambda$=457~nm,
0.01$<$intensity $I <$2~MW/cm$^{2}$). A
microscope objective of numerical aperture 0.65 (x40)
leads to a theoretical lateral resolution of approximately
300~nm. The detection pin-hole has a radius $r_{pin}$ of 10~$\mu$m.
Our ensemble is then defined by the $\simeq$200000
immobile molecules present in a detection volume of 0.5~$\mu
m^{3}$. The emitted fluorescence $F$ is detected by a
photo-multiplier in current mode coupled to a lock-in amplifier.
Let us stress that, in order to have reliable results, the
ensemble of excited molecules must not be renewed during the whole
observation time. We have tested the long-term mechanical
stability of our setup and verified that it is of the order of the
lateral resolution of the microscope.

\begin{figure}[h]
\centerline{\scalebox{0.5}{\includegraphics{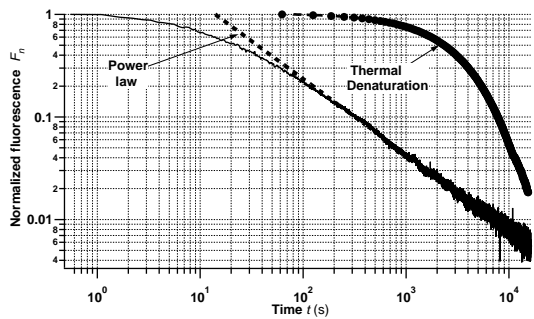}}}
  \caption{Normalized fluorescence intensity $F(t)/F(t\rightarrow 0)$ (solid line) emitted by a well defined ensemble of GFPuv ($\sim 200,000$ proteins) as a function of the total illumination time $t$. Experimental conditions: $I$=0.96~MW/cm$^{2}$ and $T$=290~K. In dotted line the power law used to
fit the long time tail: $({\tau_A\over t})^{\alpha}$ with
$\tau_A=6.8$~s and $\alpha=0.73$. For comparison we report also a
thermal aging curve obtained at $T=353$~K (circle) in a
fluorimeter together with a dotted-line curve which represents its
best exponential fit.}
\end{figure}

In Fig. 1 we present the observed time evolution of $F$ as a
function of the total illumination time $t$. All the ensembles
that we have studied under different illumination conditions
display a time evolution of $F(t)$ with a long-time tail well
described by a power law $({\tau_A\over t})^{\alpha}$
characterized by an exponent $\alpha<1$. Let us stress that in
order to unambiguously identify the asymptotical power law
behavior of $F(t)$ it is compulsory to scan several decades in
time with a good detector dynamics. The aging described by the
curve $F(t)$ appears to be an irreversible phenomenon that depends
on the illumination history of the ensemble. To investigate this
phenomenon, we switched-off the illumination at time $t_0$ and
kept the sample in the dark during a time $T_{dark}$. By
switching-on again the illumination, $F$ recovered its value at
$F(t_0)$ for $T_{dark}$ up to 24 hours. We conclude that the
unique relevant parameter for the description of the aging curves
is the total illumination time $t$. These observations show that
the transition between the fluorescent and the permanent dark
state takes place in the electronic excited state of the protein.
One may argue that the observed aging curve is the consequence of
the thermal denaturation of the protein induced by laser-heating
\cite{Penna03}. We have excluded this hypothesis by performing
several measurements of thermal denaturation in a
temperature-controlled cuvette under low intensity illumination
(uv lamp in a fluorimeter). A significant thermal denaturation of
GFPuv  is clearly observed for temperatures higher than 340~K, but
this phenomenon induces a purely exponential loss of the
fluorescence as a function of time (see Fig.~1).

In order to describe the experimental results, we propose a model that
postulates the presence of L\' evy statistics in the irreversible transition between the
fluorescent and the dark state of the protein.
Within this model,  photobleaching is the consequence of the thermally activated crossing of
a barrier in the electronic excited state of the protein (see scheme
 in Fig. 2).
\begin{figure}[h]
\centerline{\scalebox{0.45}{\includegraphics{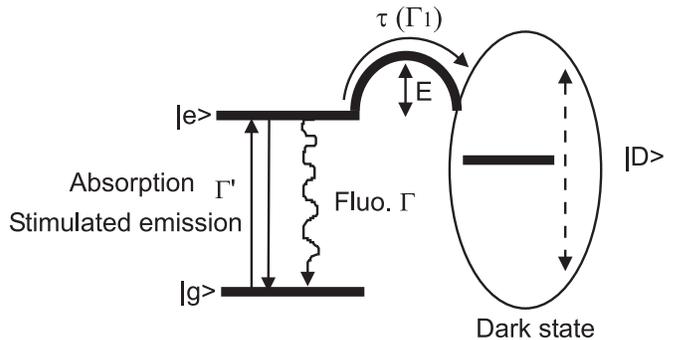}}}
 \caption{Scheme of the relevant energy-levels and coupling constants used to model the aging mechanism of the GFPuv.  $\left| g\right>$ and $\left| e\right>$ are the electronic ground and excited states of the protein, while $\left| D\right>$ represents a permanent dark state.  $\left| e\right>$ and $\left| D\right>$ are separated by a thermally-activated barrier of height $E$. $\Gamma$, $\Gamma^\prime$ and $\Gamma_1={1\over\tau}$ are, respectively, the spontaneous emission, the absorption (and stimulated emission), and the conversion towards the dark state rates.}
\end{figure}
We make the hypothesis that the heights of the barriers in a
sample are distributed according to an exponential law
$P(E)={1\over E_0}e^{-{E\over E_0}}$, where $E_0$ is the
characteristic height of the barriers. The time $\tau_i$
associated to the crossing of the $i^{th}$ barrier of height $E_i$
is then given by the Arrhenius law $\tau_i=\tau_0e^{E_i\over
k_BT}$, where $\tau_0$ is a characteristic time, $k_B$ the
Boltzmann constant and $T$ the temperature. By considering that
$P(E)dE=P(\tau)d\tau$, it is easy to show that, for $t  >>
\tau_0$,
\begin{equation}\label{dist_tau}
P(\tau)=  \alpha{\tau_0^{\alpha}\over \tau^{1+\alpha}},
\end{equation}
with $\alpha={k_BT\over E_0}$ \cite{Bardou02}. With these
assumptions, the asymptotic fluorescence at time $t$ is simply
proportional to the number of proteins that have a bright state
lifetime (time spent by the protein performing
absorption/spontaneous emission cycles) greater than the time $t$.
The fluorescence $F$ is thus proportional to the survival
probability of the molecules in the fluorescent state:
$F(t)\propto \int_t^{\infty}P(\tau)d\tau$. At long times we expect
that $F(t)\propto ({\tau_0\over t})^{\alpha}$, which exactly
corresponds to the experimental observations. Within this model,
the fact that the observed $\alpha$ coefficient is lower than 1
has a fundamental consequence: the existence of an unstable bright
state with infinite mean lifetime. The bright state is unstable in
that, at long times, each molecule ends up in a permanent dark
state with a probability tending to one. The mean lifetime is
infinite in that the ensemble average of the time $\tau$ (time of
jump to the dark state or photobleaching time) is infinite.

Let us develop now the model more in detail. The three relevant
states (scheme in Fig.~2) are coupled by the absorption and the
stimulated emission (rate $\Gamma^{\prime} \propto I $), the
spontaneous emission (rate $\Gamma$), and the bleaching process
which ends in the dark state $\left| D \right>$ (rate
$\Gamma_1={1\over\tau}$). By fixing the excitation intensity $I$ (
$\Gamma^{\prime}$) and the rate $ \Gamma_1$, we can solve a system
of coupled differential rate equations that gives us the
time-dependent population $\pi_e(\Gamma_1, I, t)$ in the
$\left|e\right>$ state. As long as we are not interested in the
short-time behavior (fast equilibration of electronic populations)
we obtain, for  $\Gamma_1<<(\Gamma+2\Gamma^\prime)$, the
simplified expression:
\begin{equation}\label{simp_pop}
\pi_e(\tau,t) \simeq {\Gamma^\prime\over\Gamma+2\Gamma^\prime} {e^{-{\Gamma^\prime\over\Gamma+2\Gamma^\prime}{t\over\tau}}}.
\end{equation}
In order to compare the model with the experiments, two fundamental ingredients have to be added to it: the spatial distribution of the excitation intensity (typically  a gaussian laser spot with characteristic dimension $r_0$
$I=I(r)=I_0e^{-({r\over r_0})^2}$) and the stochasticity associated to the rate $\Gamma_1$.

The first aspect has been discussed by A.~J.~Berglund in a theoretical paper \cite{Berglund04} that addresses the case of the samples illuminated by intensity profiles which span an infinite object plane. In this case, a gaussian laser spot and a simple model for the photobleaching (e.g. $\Gamma_1$ rate constant) lead to a time evolution of  $F$ with a long time tail that decays according to the power law $F(t)\propto {1\over t}$ \cite{Berglund04}. At first sight, this result seems to apply to our experimental observations. However in our experiment (and, more generally, in the experiments performed in a confocal geometry) the detection volume is limited by the pin-hole aperture: the pertinent intensity distribution is therefore a truncated gaussian. The consequence of this truncation is that, for $\Gamma_1$ distributed according to a narrow distribution law, $F(t)$ displays an asymptotic exponential behavior. We have checked by numerical integration of equation (\ref{simp_pop}) that, for realistic sizes of detection pin-holes (i.e. $ r_0 / 2 < r_{pin}< 2 r_0)$, the spatial distribution of the excitation intensity can only influence the details of the curve $F(t)$ at intermediate times but not its asymptotic behavior.

The properties of the statistical distribution of $\Gamma_1$ seem therefore to be the key element to understand the observed behavior.
In order to introduce this stochasticity, we have to consider the ensemble average of (\ref{simp_pop}):
$\left<\pi_e(t)\right>_{\Gamma_1}=\int \pi_e(\Gamma_1={1/\tau},t) P(\tau)d\tau$.
By introducing the  power-law distribution (\ref{dist_tau})  we obtain:
\begin{equation}
\nonumber
\left<\pi_e(t)\right>_{\tau}=
\alpha\left({\Gamma+2\Gamma'\over\Gamma'}\right)^{\alpha-1}\hskip-0.5cm\gamma\left(\alpha,{\Gamma'\over\Gamma+2\Gamma'}{t\over\tau_0}\right)\left({\tau_0\over
    t}\right)^{\alpha}
\end{equation}
where $\gamma(\alpha,x)=\int_0^x u^{\alpha-1}e^{-u} du$ is the lower incomplete gamma function.
For $t\rightarrow\infty$
this expression  becomes:
\begin{equation}\label{final_asympt}
    \left<\pi_e(t)\right>_{\tau}\simeq\alpha\left({\Gamma+2\Gamma'\over\Gamma'}\right)^{\alpha-1}\Gamma(\alpha)\left({\tau_0\over
    t}\right)^{\alpha}
\end{equation}
where $\Gamma(\alpha)$ is the gamma function.
Therefore, with the assumption of a power law distribution (\ref{dist_tau}) for the bleaching time $\tau$, the fluorescence $F(t)$ emitted by a well-defined ensemble of proteins decays asymptotically according to a power law ($\propto{1\over t^{\alpha}}$). As mentioned before, this behavior can be understood by considering that at time $t$ the term ${1\over t^{\alpha}}$ represents the survival probability of the system in the fluorescent state.
This same argument of survival probability allows us to obtain, from the measured signal $F(t)$, precious information about the distribution function of the barriers $P(E)$ without the explicit hypothesis of an exponential law.
Let us write $F(t)$ in terms of an ensemble average of the conditional probability $P_r(\tau>t|E)$ to have a lifetime $\tau>t$ with a barrier of height $E$:
\begin{equation}\label{general_E}
    F(t)=F_0\int_{0}^{\infty}P(E)dEP_r(\tau>t|E);
\end{equation}
(we note $F_0$ for $F(t\rightarrow 0)$).
The expression (\ref{general_E}) can be analytically evaluated if we replace the Poisson process associated to the Arrhenius law with a deterministic process \cite{footnote2}. In this case  $P_r(\tau>t|E)\propto 1-\Theta(t-\tau(E))$ with $\Theta$ noting the Heaviside function, and (\ref{general_E}) becomes
\begin{equation}\label{inverseE}
    F(t)=F_0\int_{E(t)}^{\infty}\hskip-0.5cm P(E)dE=F_0\int_{k_BT\ln(t/\tau_0)}^{\infty}\hskip-1.5cmP(E)dE.
\end{equation}
This expression can be easily inverted:
 \begin{equation}\label{inversion}
    P(E)=-{t\over F_0}{\partial F(t)\over\partial t}{1\over k_BT},
\end{equation}
where, again,  $E(t)$ is such that $t=\tau_0\exp(E/k_BT)$.
\begin{figure}[h]
\centerline{\scalebox{0.5}{\includegraphics{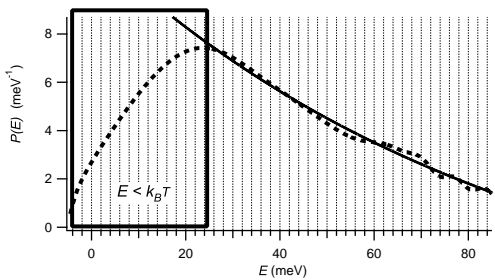}}}
  \caption{Barrier height distribution (dotted line) deduced from the experimental signal of Fig.~1 by using the expression (\ref{inversion}). The energy scale on the horizontal axis is deduced in the absence of  any free parameter. The continuous line represents the best exponential fit in the range $E>k_BT$. }
\end{figure}
We can apply such a mathematical transformation to the
experimental data of Fig.~1 and the result is represented in
Fig.~3. For energies $E>k_BT$, the distribution deduced from the
signal (dotted line) is well approximated by an exponential law
(continuous line). This result strengthens the hypothesis of an
exponential distribution of the barrier heights which was
previously introduced as a postulate. For energies $E<k_BT$ the
retrieved distribution does not follow an exponential law. This
behavior is not surprising because the Arrhenius law is no more
valid for $\tau< \tau_0$ (i.e. for $E<k_BT$ ). Let us stress that
the horizontal energy scale in Fig.~3 is obtained in the absence
of free parameters. Moreover, the order of magnitude that we
obtain for the energy barrier heights is in the typical range
expected for conformational changes of biomolecules.

\begin{figure}[h]
\centerline{\scalebox{0.5}{\includegraphics{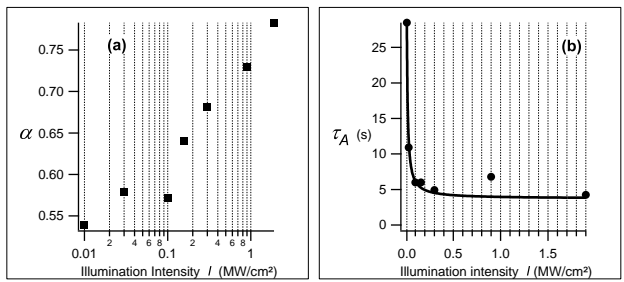}}}
  \caption{Parameters $\alpha$ (a) and $\tau_A$ (b) as a function of the excitation intensity $I$. Data are obtained by several experimental runs on different GFPuv ensembles. While the logarithmic increase of $\alpha$ is not explained by the theory, the pre-factor $\tau_A$ (filled circles in b))  follows an inverse saturation law $A+{B/ I}$ as expected from expression (\ref{final_asympt}). The continuous line represents the best fit to this law ($A=3.7$~s and $B=0.24$~s$\cdot$MW$\cdot$cm$^{-2}$).}
\end{figure}

Let us finally analyze the experimental results obtained for
several excitation intensities $I$ with different ensembles of
proteins. In order to eliminate the possible local fluctuations of
GFP concentration, we consider the normalized quantity
$F_n(t)=F(t)/F_0$. As previously mentioned, for each $I$ the
asymptotic time evolution of $F_n(t)$ is well described by an
empirical law $({\tau_A/ t})^{\alpha}$. The parameters $\tau_A$
and $\alpha$, measured as a function of $I$, are reported in Fig.
4. The exponent $\alpha$ increases slightly (with a logarithmic
law) with  $I$ (Fig. 4a). The pre-factor $\tau_A$ (filled circles
in Fig. 4b) follows an inverse saturation law $A+{B/ I}$ (the
continuous line in Fig. 4b is the best fit to this law). The
logarithmic behavior of $\alpha$ is not explained by our model in
which  $\alpha$ only depends on the temperature. On the contrary,
the expression (\ref{final_asympt}) accounts for the observed $I$
dependency of the pre-factor $\tau_A$. Indeed, by considering the
normalized excited state population
${\left<\pi_e(t)\right>_{\tau}\over\pi_e(t\rightarrow
0)}={\Gamma+2\Gamma^\prime\over\Gamma^\prime}\left<\pi_e(t)\right>_{\tau}$,
we obtain from expression (\ref{final_asympt}) that
$\tau_A^\alpha=\alpha\left(\tau_0{\Gamma+2\Gamma'\over\Gamma'}\right)^{\alpha}\Gamma(\alpha)$.
By considering that $\Gamma^\prime\propto I$, we obtain the
empirical law that we used in Fig.~4b to fit the experimental
data.

In conclusion, we have found experimental evidence of the fact that the photobleaching of a
well-defined ensemble of Green Fluorescent Proteins is governed by L\'evy statistics.
The model that we have developed is based on an irreversible thermally-activated conversion between an excited electronic state and a dark state.
This model allows us to explain the observed temporal decay of the fluorescence. Moreover, it gives us quantitative information about the probability distribution of the barrier heights involved in the aging mechanism.
Within this simple model the exponent $\alpha$ should only depend on temperature. Further experiments, beyond the scope of this paper, will investigate this $T$ dependence and the observed logarithmic $I$ dependence of the $\alpha$ coefficient.
\begin{acknowledgments}
We thanks C.~Brochon, F.~Linker and E. Weiss for their help in samples preparation; J.-Y.~Bigot and O.~Cr\'egut for useful discussions; M.~Albrecht for technical support.
\end{acknowledgments}

\begin{thebibliography}{30}
\bibitem{footnote2}
    Without this hypothesis, Eq.~\ref{general_E} can be expressed in terms of a truncated Laplace Transform.
\end{thebibliography}
\end{document}